\documentclass[reprint,superscriptaddress,floatfix]{revtex4-2}

\usepackage{graphicx}
\usepackage{float}
\usepackage{svg}
\usepackage[mode=tex]{standalone}
\usepackage{tikz}
\usetikzlibrary{shapes}
\usetikzlibrary{positioning}
\usetikzlibrary{calc}
\usetikzlibrary{fit}
\usetikzlibrary{arrows.meta, shapes.misc}

\usepackage{xparse}

\definecolor{c0}{HTML}{1f77b4}
\definecolor{c1}{HTML}{ff7f0e}
\definecolor{c2}{HTML}{2ca02c}
\definecolor{c3}{HTML}{d62728}
\definecolor{c4}{HTML}{9467bd}
\definecolor{c5}{HTML}{8c564b}
\definecolor{c6}{HTML}{e377c2}
\definecolor{c7}{HTML}{7f7f7f}
\definecolor{c8}{HTML}{bcbd22}
\definecolor{c9}{HTML}{17becf}

\tikzset{
dot/.style = {circle, fill=black, minimum size=#1,
              inner sep=0pt, outer sep=0pt},
dot/.default = 6pt 
}

\usepackage{dcolumn}
\usepackage{bm}
\usepackage{bbold}
\definecolor{linkred}{RGB}{160,0,0}
\definecolor{citegreen}{RGB}{0,160,0}
\definecolor{urlblue}{RGB}{0,0,160}
\usepackage[pdfstartview=FitH,      
            breaklinks=true,        
            bookmarksopen=true,     
            bookmarksnumbered=true, 
            colorlinks=true, urlcolor=urlblue, pdfborder={0 0 0},
            linkcolor=linkred,
            citecolor=citegreen,
            pdfencoding=auto,
            psdextra
            ]{hyperref}

\usepackage{braket}
\usepackage[obeyFinal,section]{easy-todo}
\usepackage{cleveref}

\usepackage[ruled,vlined]{algorithm2e}
\SetKwComment{Comment}{$\triangleright$\ }{}

\SetCommentSty{mycommfont}

\usepackage{pgfplots}
\graphicspath{{figures/}{.}}
\makeatletter
\def\input@path{{figures/}{.}}
\makeatother

\pgfplotsset{compat=1.16}
\providecommand\inputpgf[2]{{
\let\pgfimageWithoutPath\pgfimage
\renewcommand{\pgfimage}[2][]{\pgfimageWithoutPath[##1]{#1/##2}}
\input{#1/#2}
}}

\begin{document}

\preprint{APS/123-QED}

\title{Probing ground state properties of the kagome antiferromagnetic
       Heisenberg model using the Variational Quantum Eigensolver}

\author{Jan Lukas Bosse}
  \email{janlukas.bosse@bristol.ac.uk}
\affiliation{School of Mathematics, University of Bristol}
  \affiliation{Phasecraft Ltd.}
\author{Ashley Montanaro}%
 \email{ashley@phasecraft.io}
\affiliation{Phasecraft Ltd.}
\affiliation{School of Mathematics, University of Bristol}

%
%

\date{\today}

\begin{abstract}
  Finding and probing the ground states of spin lattices, such as the
  antiferromagnetic Heisenberg model on the kagome lattice (KAFH), is a very
  challenging problem on classical computers and only possible for relatively
  small systems. We propose using the Variational Quantum Eigensolver (VQE) to
  find the ground state of the KAFH on a quantum computer.  We find efficient
  ansatz circuits and show how physically interesting observables can be
  measured efficiently.  To investigate the expressiveness and scaling of our
  ansatz circuits we used classical, exact simulations of VQE for the KAFH for
  different lattices ranging from 8 to 24 qubits.  We find that the fidelity
  with the ground state approaches one exponentially in the circuit depth for
  all lattices considered, except for a 24-qubit lattice with an almost
  degenerate ground state.  We conclude that VQE circuits that are able to
  represent the ground state of the KAFH on lattices inaccessible to exact
  diagonalisation techniques may be achievable on near term quantum hardware.
  However, for large systems circuits with many variational parameters are
  needed to achieve high fidelity with the ground state.
\end{abstract}

\maketitle


Simulating and investigating quantum mechanical systems will be one of the first
applications of the noisy intermediate-scale quantum (NISQ) computing hardware
available in the near term~\cite{Cirac_2012,Preskill_2018,Barthi_2021}. In particular
problems in quantum many-body physics and quantum chemistry are hard to 
intractable for today's best supercomputers and of high scientific interest.

One algorithm that was devised with the low circuit depths and relatively high
noise rates of NISQ devices in mind is the Variational Quantum Eigensolver
(VQE)~\cite{Peruzzo2014}. It is a hybrid quantum-classical algorithm to produce
a ground state of a quantum Hamiltonian $H$. A classical optimiser is used to
minimise the expectation value $\braket{\psi(\theta) |H| \psi(\theta)}$ over a
family of states $\ket{\psi(\theta)}$. If this family of states includes a
ground state of $H$ VQE aims to return this ground state.

In this work we focus on a quantum many-body system that is hard to study using
classical supercomputers and that we hope is particularly suitable to be
addressed using VQE on NISQ quantum computers: The antiferromagnetic Heisenberg
model on the kagome lattice (KAFH). The phase of its ground state is still not
entirely settled. Some analytical results indicate a valence bond crystal
ground state~\cite{Nikolic_2003,Huse_2007} while other results indicate an
algebraic spin liquid ground state~\cite{Wen_2007,Wen_2008}. More recent
numerical results predict the ground state to be a
gapped~\cite{Schollwoeck_2012,Yan_2011,Mei_2017} or
gapless~\cite{Iqbal_2013,Liao_2017,Jiang_2008} spin liquid state.
Experimental results with herbertsmithite ZnCu${}_3$(OH)${}_6$Cl${}_2$ also
indicate a gapless spin liquid state~\cite{Lee_2015,Khuntia_2020}. The
classical, numerical methods are limited in their ability to describe large
systems (exact diagonalisation (ED)) or to describe true 2D systems instead of
mimicking them using long strips or cylinders (DMRG). In contrast, the KAFH
maps well to NISQ quantum hardware, because the qubits can directly represent
the spin-$\frac{1}{2}$ degrees of freedom and many NISQ devices have a 2D
connectivity close to that of the kagome lattice.

Here we develop efficient quantum circuits for VQE for KAFH targeted at
nearest-neighbour quantum computing architectures, based on the popular and
physically-motivated Hamiltonian Variational (HV) ansatz~\cite{Wecker2015}. We
then carry out exact, classical simulations of VQE for the KAFH with the HV
ansatz on patches of the kagome lattice from 8 to 24 sites using variants of
the HV ansatz with different levels of parametrisation. These extensive
experiments allow us to infer the possible scaling of VQE for larger system
sizes beyond the capability of classical exact diagonalisation. We also show
that some physically interesting observables in the states produced by VQE
match their values in the exact ground state.

We find that, similarly to previous work on VQE for the Hubbard
model~\cite{Cade_2019} and interacting, spinless fermions on a
chain~\cite{Biamonte_2020}, the error as measured by infidelity decreases
exponentially with the VQE circuit depth.  Using one ansatz parametrisation,
our results are largely consistent with ground state fidelity 0.999 being
achievable using roughly half the number of qubits,  $0.5 {n_\mathrm{qb}}$, VQE
layers for strip-shaped lattices.  Although we have insufficient data for
lattices with equal length in both dimensions to determine scaling, consideration of entanglement spreading out linearly in the circuit depth suggests the
scaling could even be as low as $\sqrt{n_{\mathrm{qb}}}$.  We show that each
VQE layer can be implemented using a quantum circuit with 2-qubit gate depth 7
on a quantum computer with square lattice connectivity, or only 4 on an
architecture with all-to-all connectivity. 

However, this performance and the fastest exponential decay in infidelity is
only achieved by a variant of the HV ansatz with one independent parameter per
edge in the lattice and VQE layer ($\approx 2pn_{\mathrm{qb}}$ parameters for a
system of $n_{\mathrm{qb}}$ qubits and $p$ VQE layers). The use of VQE with a
large number of parameters may be expected to suffer from the ``barren
plateau'' problem~\cite{McClean_2018,Holmes_2021} of exponentially small
gradients, and indeed we do find some evidence that these become small for
large $p$.  Also, in one case (a patch of the kagome lattice on 24 sites), VQE
does not find the ground state in a reasonable number of layers. This is
because the first excited state is very close in energy to the ground state; we
find additionally that the overlap achieved with the subspace spanned by these
two states is significantly better.

It is instructive to compare the complexity of VQE for KAFH with VQE for the
Fermi-Hubbard model, which is another plausible early application for NISQ
quantum computers, and was studied in detail in~\cite{Cade_2019,Cai2020}. It
was argued in~\cite{Cade_2019} that a $5\times 5$ Fermi-Hubbard instance (50
spins) might be solvable with high accuracy via a VQE circuit with overall
two-qubit gate depth $\approx 550$ in a square-lattice topology, which was
substantially lower than other proposed near-term applications of quantum
computers. If our numerical results for KAFH are representative of the
performance of VQE for larger systems, a patch of the kagome lattice with 50
qubits could be solved with two-qubit gate depth only $\approx 0.5 \times 50
\times 7 = 175$ in the same topology, which is substantially lower. However,
the variational ansatz used in~\cite{Cade_2019} would use 125 parameters for
the whole circuit, whereas the ansatz we use here would use 2500 parameters.

These points highlight the opportunities and challenges faced by VQE for
condensed-matter systems. For small-scale versions of complex systems of
significant physical interest, VQE ans\"atze can well-represent the ground
state and can be implemented efficiently. Yet the VQE algorithm may struggle
with finding the ground state where there is a small energy gap, and with
handling optimisation over a large number of variational parameters.

During the completion of this project we became aware of closely related work
done by Kattemölle and van Wezel~\cite{Kattemoelle_2021}. They also study the
performance of VQE with the HV ansatz on the KAFH, including numerical
simulations on 20 qubits. However, they consider a different embedding of the
kagome lattice into the square lattice that produces one round of the HV ansatz
with one layer of two-qubit gates less than ours at the cost of one extra qubit
per unit cell of the kagome lattice. Consequently, their ansatz circuits are
different from ours and, unlike ours, fully respect the translational symmetry
of the kagome lattice. They report slightly better scaling of the fidelity with
circuit depth for a lattice on 20 qubits, but more detailed comparisons on more
lattices are needed to draw any definitive conclusions regarding scaling.
Other differences are the inclusion in this work of the study of gradient
scaling---to understand whether barren plateaus are a problem in our ansatz
circuits---and comparison of local observables for VQE states and exact ground
states---to understand how well VQE states represent ground state properties
other than the energy.

\section{The variational method}
\label{sec:the-variational-method}

The Variational Quantum Eigensolver (VQE)~\cite{Peruzzo2014, McClean_2016} is a
method for finding ground states (or possibly also excited states
\cite{Higgott_2019}) of quantum Hamiltonians by classically optimising the
parameters of a parametric unitary $U(\theta)$. It has been extensively studied
using classical simulation~\cite{Cade_2019,Biamonte_2020,Mohtashim_2021} as
well as on superconducting quantum computers~\cite{Kandala_2017,GoogleVQE_2020,Ganzhorn_2019} and with trapped ions~\cite{Hempel_2018,Shen_2017,Nam_2019}.

At a high level, the goal is to minimize the objective function
\begin{equation}
  f(\theta) = \braket{\psi(\theta) | H | \psi(\theta)}
       =\braket{\psi_i| U^\dagger(\theta) H U(\theta) |\psi_i},
\end{equation}
where $U(\theta)$ is a family of unitaries parametrised by the classical
parameters $\theta$, $\ket{\psi_i}$ an easily prepared reference state, and
$H$ the Hamiltonian whose ground state $\ket{\psi_0}$ we wish to prepare. If and only if
$U(\theta)$ is sufficiently expressive~\cite{Hubregtsen_2021,Holmes_2021}
(meaning there exists some $\theta^*$ such that $\ket{\psi(\theta^*)} =
\ket{\psi_0}$), then $f(\theta)$ is minimal if and only if $\ket{\psi(\theta)}$
is the ground state  of $H$.

To turn this high level description into a concrete quantum algorithm, we need
to specify its components: We need to encode the system of interest, and hence
the Hamiltonian $H$, into a qubit system, we need to make some (informed)
choice of a sufficiently expressive yet implementable ansatz circuit
$U(\theta)$, and we need a way to efficiently estimate the expectation value
$\braket{\psi(\theta) | H | \psi(\theta)}$ from measurement samples. Unless the
goal of the experiment is to find the ground state energy $E_0$, we also need
to specify what to do with the found ground state, e.g.\ measuring some
observables of interest.

\subsection{Encoding and lattice mapping}
\label{sec:encoding-and-lattice-mapping}

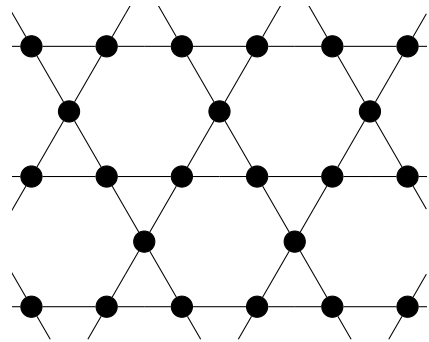
\begin{figure}
  \centering
  \newcommand{\kagomeunitcell}[1]{
  \begin{scope}[shift={#1}]
    \node [draw, shape=circle, fill=black, inner sep=0pt, minimum size=8pt] (0) at (-0.5, {.5*sqrt(3)}){};
    \node [draw, shape=circle, fill=black, inner sep=0pt, minimum size=8pt] (1) at (0,0){};
    \node [draw, shape=circle, fill=black, inner sep=0pt, minimum size=8pt] (2) at (1,0){};
    \draw (0) -- (1);
    \draw (1) -- (2);
    \draw (0) -- ++(-.25, {-.25*sqrt(3)});
    \draw (0) -- ++(+.25, {+.25*sqrt(3)});
    \draw (0) -- ++(-.25, {+.25*sqrt(3)});
    \draw (1) -- ++(-.5, 0);
    \draw (1) -- ++(+.25, {-.25*sqrt(3)});
    \draw (2) -- ++(+.5, 0);
    \draw (2) -- ++(-.25, {-.25*sqrt(3)});
    \draw (2) -- ++(+.25, {+.25*sqrt(3)});
  \end{scope}
}

\begin{tikzpicture}
  \clip (-.25,-.5) rectangle (5.25,4);

  \kagomeunitcell{(-2,0)}
  \kagomeunitcell{(0,0)}
  \kagomeunitcell{(2,0)}
  \kagomeunitcell{(4,0)}
  \kagomeunitcell{(-3,{sqrt(3)})}
  \kagomeunitcell{(-1,{sqrt(3)})}
  \kagomeunitcell{(1,{sqrt(3)})}
  \kagomeunitcell{(3,{sqrt(3)})}
  \kagomeunitcell{(5,{sqrt(3)})}
  \kagomeunitcell{(-2,{2*sqrt(3)})}
  \kagomeunitcell{(0,{2*sqrt(3)})}
  \kagomeunitcell{(2,{2*sqrt(3)})}
  \kagomeunitcell{(4,{2*sqrt(3)})}

\end{tikzpicture}
  \caption{The kagome lattice}
  \label{fig:kagome_lattice}
\end{figure}

Our target Hamiltonian is the kagome antiferromagnetic Heisenberg model (KAFH)
\begin{equation}
  H = \sum_{\braket{i,j}} \vec{S}_i \cdot \vec{S}_j
  = \sum_{\braket{i,j}} X_i X_j + Y_i Y_j + Z_i Z_j,
  \label{eq:kafh_hamiltonian}
\end{equation}
where the sums run over neighbouring sites $\braket{i,j}$ in the kagome lattice
(\cref{fig:kagome_lattice}), $\hbar = 1$ and the usual factor of $\frac{1}{2}$ in 
the definition of the spin operators $\vec{S}_i$ is omitted for simplicity.

As a spin model, the Heisenberg model \cref{eq:kafh_hamiltonian} has a
canonical mapping to qubits: Simply assign one spin to each qubit. Finding an
efficient mapping from the kagome to the square lattice is more challenging.
The HV ansatz requires time-evolution generated by the terms in the Hamiltonian.
Hence, we seek a mapping that can implement a trotterised version of $e^{-itH}$
---that is, to implement time-evolution by all of the terms in $H$
individually---with as few layers of parallel two-qubit gates as possible on a
square lattice. The chosen mapping is
shown in \cref{fig:kagome_lattice_square} and with it a trotterised version of
$e^{-i t H}$ can be implemented in 7 layers of two-qubit gates on a
nearest-neighbour architecture, as shown in
\cref{fig:kagome_lattice_square_list}. In a system with all-to-all
connectivity, such as certain ion trap architectures, the two-qubit gate depth
can even be reduced to 4 layers. This is because the edges of the kagome
lattice can be coloured with 4 colours such that no two edges with the same
colour share a vertex, as shown in \cref{sec:a-4-colouring-of-the-kagome-lattice}.
Then all terms in $H$ corresponding to the same colour can be implemented in
parallel. Kattemölle and van Wezel~\cite{Kattemoelle_2021} implemented 
this ansatz for periodic lattices on 12, 18 and 24 qubits and found results
comparable to ours shown in \cref{fig:lattice_comparison}.

\begin{figure*}[htp]
\centering
  \begin{minipage}[c]{.49\textwidth}
    \includestandalone{kagome-lattice-square} 
    \caption{
    Mapping the kagome lattice onto a square lattice. The colours of the lines
    show which interactions exist only in the kagome lattice, only in the square
    lattice and which exist in both. The line types show which pairs of qubits 
    must be swapped to implement the interactions existing in the kagome lattice
    but missing in the square lattice while the numbers group the interactions 
    that can be implemented in parallel.
    }
    \label{fig:kagome_lattice_square}
  \end{minipage} \hfill
  \begin{minipage}[c]{.49\textwidth}
    \begin{enumerate}
      \item Let all \tikz[baseline=-.5ex]{\draw[c0] (0,0) -- (0.7,0)
                      node[midway, inner sep=0, fill=white](){1}},
                    \tikz[baseline=-.5ex]{\draw[c0,<->] (0,0.1) -- (0.7,-0.1)
                      node[midway, inner sep=0, fill=white](){1}} and 
                    \tikz[baseline=-.5ex]{\draw[c0,double] (0,-0.1) -- (0.7,0.1)
                      node[midway, inner sep=0, fill=white](){1}}
            connected qubits interact
      \item Let all \tikz[baseline=-.5ex]{\draw[c0] (0,0) -- (0.7,0)
                      node[midway, inner sep=0, fill=white](){2}},
                    \tikz[baseline=-.5ex]{\draw[c0,<->] (0,0.1) -- (0.7,-0.1)
                      node[midway, inner sep=0, fill=white](){2}} and
                    \tikz[baseline=-.5ex]{\draw[c0,double] (0,-0.1) -- (0.7,0.1)
                      node[midway, inner sep=0, fill=white](){2}}
                connected qubits interact
      \item Let all \tikz[baseline=-.5ex]{\draw[c0] (0,0.2) -- (0.2,-0.2)
                      node[midway, inner sep=0, fill=white](){3}}
            connected qubits interact
      \item Swap all \tikz[baseline=-.5ex]{\draw[c0,<->] (0,0.1) -- (0.7,-0.1)
                      node[midway, inner sep=0, fill=white](){1}} and 
                     \tikz[baseline=-.5ex]{\draw[c0,<->] (0,0.1) -- (0.7,-0.1)
                      node[midway, inner sep=0, fill=white](){2}} 
            connected pairs of qubits
      \item Let \tikz[baseline=-.5ex]{\draw[c1] (0,-0.2) -- (0.4, 0.2)
                      node[midway, inner sep=0, fill=white](){4}} 
            connected qubits interact by 
                \tikz[baseline=-.5ex]{\draw[gray,double] (0,0.2) -- (0.2,-0.2)
                      node[midway, inner sep=0, fill=white](){}}
      \item Let \tikz[baseline=-.5ex]{\draw[c1] (0,0) -- (0.7, 0)
                      node[midway, inner sep=0, fill=white](){5}} 
            connected qubits interact by 
                \tikz[baseline=-.5ex]{\draw[c0,double] (0,-0.1) -- (0.7, 0.1)
                      node[midway, inner sep=0, fill=white](){1/2}}
      \item Swap \tikz[baseline=-.5ex]{\draw[c0,<->] (0,0.1) -- (0.7,-0.1)
                      node[midway, inner sep=0, fill=white](){1}} and 
                     \tikz[baseline=-.5ex]{\draw[c0,<->] (0,0.1) -- (0.7,-0.1)
                      node[midway, inner sep=0, fill=white](){2}} 
            back for initial configuration
    \end{enumerate}
    \caption{
    One full round of 2-qubit interactions needed to implement all interactions
    of the kagome lattice in the square lattice. Here we usually assume that any
    2-qubit gate can be implemented as one elementary operation. With a
    $\sqrt{i\mathrm{SWAP}}$ native gate, as used in some architectures, the
    interaction gates take two 2-qubit gates and the SWAPs take three 2-qubit
    gates.
    }
    \label{fig:kagome_lattice_square_list}
  \end{minipage}
\end{figure*}

\subsection{Ansatz circuits}
\label{sec:ansatz-circuits}
The barren plateau problem~\cite{McClean_2018,Holmes_2021} shows that for VQE
to be practical even at larger system sizes or for deeper circuits it is
imperative to make use of prior knowledge about the problem in the ansatz
circuits $U(\theta)$. One such ansatz is the Hamiltonian Variational (HV)
ansatz~\cite{Wecker2015}. It is based on the adiabatic theorem of quantum
mechanics: If a system is prepared in the ground state of an (simple) initial
Hamiltonian $H_0$ and then evolved under a time-dependent Hamiltonian $H(t)$,
changing sufficiently slowly from $H_0$ into a target Hamiltonian $H$, the
system will remain in the ground state and we end up with the ground state of
$H_1$. Setting $H(t) = (1 - \frac{t}{T}) H_0 + \frac{t}{T} H$ the time
evolution can be approximately implemented by trotterisation, i.e.
alternatingly applying $e^{-i H_0 \Delta t}$ and $e^{-i H \Delta t}$ for
short times $\Delta t$ and possibly also trotterising $e^{-i H_0 \Delta t}$ and
$e^{-i H \Delta t}$. To turn this intuition into a variational circuit the
short $\Delta t$'s are replaced by longer evolution times $\theta_i$ which are
then optimised. This makes the ansatz circuit with $2p$ layers:
\begin{equation}
  U(\theta) = \prod_{i=1}^p e^{-i \theta_{i,0} H_0} e^{-i \theta_{i,1} H}.
  \label{eq:per_hamiltonian}
\end{equation}
We call this ansatz the ``per hamiltonian'' ansatz, because it has one
parameter per application of $H_0$ and $H$.  In our case, $e^{-i t H}$ cannot
be directly implemented, because the individual terms in
\cref{eq:kafh_hamiltonian} don't commute. Instead we split $H = \sum_{j=1}^5
H_j$ according to the five subgraphs labelled in
\cref{fig:kagome_lattice_square}.  This yields an alternative parametrisation
with $6p$ parameters:
\begin{equation}
  U(\theta) = \prod_{i=1}^p \left[ e^{-i \theta_{i,0} H_0}
  \kern-1em \prod_{j\in [3,2,1,5,4]} \kern-1em e^{-i \theta_{i,j} H_j} \right]
  \label{eq:per_edge_color}
\end{equation}
where $\prod_{j \in [j_1,\cdots,j_n]} U_j = U_{j_1} \cdots U_{j_n}$ denotes the
product ordering.  Because the decomposition $H = \sum_{j=1}^5 H_j$ corresponds
to an edge colouring of the lattice we call this ansatz the ``per edge color''
ansatz. A variant of this ansatz is the ``per edge color ii'' ansatz, where we
drop the $e^{-i \theta_{i,0} H_0}$ factors because we choose $H_0 = H_1$.  The
``per edge color'' ansatz can be made even more expressive by allowing
different evolution times for all interaction terms in
\cref{eq:kafh_hamiltonian} and using
\begin{equation}
  U(\theta) = \prod_{i=1}^p \left[ e^{-i \theta_{i,0} H_0}
  \prod_{\braket{k,l}} e^{-i \theta_{i,kl} \vec{S}_k \cdot \vec{S}_l} \right].
  \label{eq:per_gate}
\end{equation}
Here, the terms in the inner product are in the same order as in
\cref{eq:per_edge_color}. Now each parameter corresponds to an edge in the
lattice, hence we dub this ansatz the ``per edge'' ansatz.  Note that the
circuits described by \cref{eq:per_hamiltonian,eq:per_edge_color,eq:per_gate}
are all the same. The ansätze only differ in the number of independent
parameters.

As an initial Hamiltonian $H_0$ we use the Heisenberg Hamiltonian on the dimer
covering induced by the 
\tikz[baseline=-.5ex]{\draw[c0] (0,0) -- (1,0) node[midway, inner sep=0, fill=white](){1}},
\tikz[baseline=-.5ex]{\draw[c0,<->] (0,0.1) -- (0.7,-0.1) node[midway, inner sep=0, fill=white](){1}}
and
\tikz[baseline=-.5ex]{\draw[c0,double] (0,-0.1) -- (0.7,0.1) node[midway, inner sep=0, fill=white](){1}}
terms. Its ground state $\ket{\psi_i}$ is the product of singlets 
$\ket{s=0} = \frac{1}{\sqrt{2}}[\ket{\downarrow \uparrow} - \ket{\uparrow \downarrow}]$ on 
all connected pairs of spins. It can be prepared  via %
\begin{equation}
  \ket{s=0}_{kl} =
  \mathrm{Z}_k \, \mathrm{X}_l \, \mathrm{CNOT}_{kl} \, \mathrm{H}_k \ket{\downarrow \downarrow}
  \label{eq:prepare_singlet_state}
\end{equation}
and for an even number of qubits it is in the same $S^x = \sum_{i} X_i = 0$, 
$S^y = \sum_{i} Y_i = 0$ and $S^z = \sum_{i} Z_i = 0$ symmetry sectors as 
the ground state of the KAFH.

All three ansätze \cref{eq:per_hamiltonian,eq:per_edge_color,eq:per_gate}
preserve all those three symmetries, meaning we stay in the correct symmetry
sector throughout the whole circuit. Additionally, this also gives a way to
post-select samples: If a sample does not satisfy $S^\alpha = 0$, we know that
an error must have occurred and we can discard that sample.

\subsection{Measurements}
\label{sec:measurements}
Reducing the number of measurements needed to estimate expectation values in 
VQE has been extensively studied~\cite{Verteletskyi2020,Verteletskyi2020_ii,Crawford2021}.
Most strategies are based on cleverly grouping the terms of the Hamiltonian 
such that all the terms in one group can be measured simultaneously, possibly
after changing the measurement basis with some simple, local unitaries.
In our case, the Hamiltonian can be decomposed as 
\begin{equation}
  H = \sum_{\braket{i,j}} X_i X_j
    + \sum_{\braket{i,j}} Y_i Y_j
    + \sum_{\braket{i,j}} Z_i Z_j
\end{equation}
and each of the three terms can be estimated separately by measuring all 
qubits in the $X$, $Y$ or $Z$-basis.

\subsection{Probing observables}
\label{sec:probing-observables}

If VQE succeeds in preparing the ground state of the KAFH, that state can be
further used to investigate the nature this ground state. Many of the
observables used in classical numerical or experimental studies are also
readily measurable on quantum computers, such as the ground state
energy or correlation functions.  Others, e.g.\ entanglement entropies
or energy gaps, are more challenging to measure on a quantum computer.

\subsubsection{Ground state and spin gap}
\label{sec:ground state-and-spin-gap}
The ground state energy is trivially measurable; $\braket{H}$ is exactly the
cost function minimized by VQE. In contrast, the energy of the first excited
state, and hence the energy gap, is not as easy to obtain. There exist several
proposals for VQE algorithms to find excited energy states of Hamiltonians.
The proposals most feasible on NISQ hardware are based on augmenting the
Hamiltonian with projectors on the $k-1$ lowest lying states to find the $k$-th
eigenstate,
\begin{equation}
  H_k = H + \sum_j^{k-1} \alpha_j \ket{\psi_j} \bra{\psi_j}
  \quad \textnormal{with} \quad
  \alpha_j > 0
  \label{eq:higher_state_vqe}
\end{equation}
such that the $k$-th eigenstate is now the ground state of $H_k$ and can be
found by the usual VQE method. However, to estimate the expectation value of
the second term in \cref{eq:higher_state_vqe} one needs to either double the
circuit depth \cite{Higgott_2019} or double the number of qubits
needed~\cite{Jones_2019}. The subspace-search VQE
algorithm~\cite{Nakanishi_2019} does away with the need of doubling the number
of qubits or circuit depth, but potentially needs more complex ansatz circuits
that are not as well motivated as the HV ansatz. Overall, none of the methods
reviewed in~\cite{Kuroiwa_2021} has as low requirements as the original VQE
problem.  Thus, if we are only just able to prepare the ground state with VQE
we will be unable to prepare excited states for the same system.

But because the Heisenberg model, as well as our ansatz circuits, conserves the 
total $S^x, S^y$ and $S^z$ spin it is possible to obtain the ground states and 
energies within each symmetry sector separately by simply choosing $\ket{\psi_i}$
to be in that symmetry sector. This makes measuring the \emph{spin gap}
\begin{equation}
    \Delta E_S = 
    \begin{cases}
        E_0(S^z=1) - E_0(S^z=0) \qquad \textrm{even } N \\
        E_0\left(S^z=\frac{3}{2}\right) - E_0\left(S^z=\frac{1}{2}\right) \qquad \textrm{odd } N
    \end{cases}
\end{equation}
as easy (or hard) as the original VQE problem. For consistency with the
literature we used the spin operators with eigenvalues $\pm \frac{1}{2}$ here,
unlike in the rest of this paper.

In principle, VQE could use coherent noise in the gates to break the spin 
conservation to get from a $S^\alpha = 1$ into a $S^\alpha = 0$ sector 
and thus lower the energy. However, post-selection of the samples in the correct
spin sectors ensures the energy cannot be lowered through this process.

\subsubsection{Correlation functions and structure factors}
\label{sec:correlations-and-structure-factors}

The presence or absence of long-range order is indicated by 
the decay of the \emph{spin-spin}
\begin{equation}
  C_S(\vec{i},\vec{j}) = \left| \braket{ S_i^z S_j^z } \right|
\end{equation}
and \emph{dimer-dimer correlations}
\begin{equation}
  C_D(\vec{ij},\vec{kl}) =
    \braket{\vec{S}_{i} \cdot \vec{S}_{j}
         \, \vec{S}_{k} \cdot \vec{S}_{l}}
      - 
    \braket{\vec{S}_{i} \cdot \vec{S}_{j}}
    \braket{\vec{S}_{k} \cdot \vec{S}_{l}},
  \label{eq:dimer-dimer-correlation}
\end{equation}
where $\vec{i}$ and $\vec{j}$ are on adjacent sites and similarly for $\vec{kl}$.
In this section $\vec{i}$ denotes the real-space location of the $i$-th qubit 
on the lattice and difference vectors are written as $\vec{ij}$.
One sample of \emph{all} spin-spin correlations can be obtained by measuring all
qubits in the computational basis once. The dimer-dimer correlation between 
two edges that share no qubit can be measured by running the reverse circuit 
of \cref{eq:prepare_singlet_state} on each of the dimers and then measuring in
the computational basis, the result $\ket{\downarrow \downarrow}_{ij}$ means
$\vec{S}_i \cdot \vec{S}_j = -3$ while the three other outcomes mean
$\vec{S}_i \cdot \vec{S}_j = +1$. Note that the first term in
\cref{eq:dimer-dimer-correlation} cannot be reconstructed from the measurements 
in the $X, Y$ and $Z$ basis that were used to estimate $\braket{H}$,
because it also contains terms of the form $X_i X_j Y_k Y_l$.

The spatial structure of the ground state is revealed by the
\emph{static spin structure factor}
\begin{equation}
  S^z(\vec{q}) = \frac{1}{N} \sum_{i,j}
		    e^{\imath \vec{q} \cdot (\vec{i} - \vec{j})}
		    \left< S_i^z S_j^z \right>,
\end{equation}
which can also be obtained from measurements of all qubits in the computational
basis.  In fact, all data needed to compute this static spin structure factor
and the spin-spin correlations is already available from the VQE optimisation;
to measure the $\sum_{\braket{i,j}}Z_i Z_j$-part of the Hamiltonian we already
needed to measure all qubits in the computational basis.

\section{Results}
\label{sec:results}

We carried out extensive numerical simulations to investigate the performance
of our ansatz circuits. We were mainly interested in whether our ansatz circuits
are expressive enough to represent the ground state, and if yes at which depth. 
Since the end goal is to determine the nature of the ground state by measuring 
different observables, we also studied how closely the states found by VQE 
represent these observables.
Because of this focus on expressibility we computed exact expectation values 
$f(\theta) = \braket{\psi(\theta)|H|\psi(\theta)}$ from the full wave functions 
and also exact, analytical gradients $\partial_{\theta^\mu} f(\theta)$ using 
\texttt{Yao.jl}'s~\cite{YaoFramework2019} automatic differentiation algorithms.
As a classical optimisation algorithm, we used the L-BFGS algorithm from the 
NLopt optimisation suite. Note that this is only a good choice for exact, 
classical simulations. On real hardware, or if the expectation value 
is computed from measurement samples instead of from the full wave function, 
it does not work, because it relies on the knowledge of exact gradients.
As a reference, we also computed the exact ground state for each Hamiltonian
using the Lanczos algorithm implemented in ARPACK from a sparse representation
of the full Hamiltonian.  The simulations were carried out on the Google Cloud
Platform using $16 \times 2.8 \, \textrm{GHz}$ Intel Xenon CPUs for the lattice
sizes $\leq 16$ qubits and on $2\times$ NVIDIA Tesla T4 GPUs for the larger
lattices.

\begin{figure}[tb]
  \centering
  \input{figures/graphs.pgf}
  \caption[The six graphs used for classical simulations]{
    The six graphs used for classical simulations. The graphs are mapped to a
    square lattice as described in
    \cref{fig:kagome_lattice_square,fig:kagome_lattice_square_list}.  Because
    we chose the dimer covering on all 
    \tikz[baseline=-.5ex]{\draw[c0] (0,0) -- (0.7,0)
                          node[midway, inner sep=0, fill=white](){1}},
    \tikz[baseline=-.5ex]{\draw[c0,<->] (0,0.1) -- (0.7,-0.1)
                          node[midway, inner sep=0, fill=white](){1}} and 
    \tikz[baseline=-.5ex]{\draw[c0,double] (0,-0.1) -- (0.7,0.1)
                          node[midway, inner sep=0, fill=white](){1}}
    connected qubits as the initial state for VQE and want to start in the 
    right $S^x, S^y$ and $S^z$ sectors we had to restrict ourselves to even 
    numbers of columns. In the $2\times$, $2\times 10$ and $3\times 8$ lattice
    we marked the qubits between which we computed the correlations in 
    \cref{sec:results-observables} with bigger nodes.
}%
  \label{fig:graphs}
\end{figure}

To study the scaling of the attainable fidelity as a function of the number of
ansatz layers $p$ and lattice size we ran classical simulations of VQE on the 6
different lattices shown in \cref{fig:graphs}, ranging in size from 8 to 24
qubits. For all simulations, we ran all three different ansatz circuits
\cref{eq:per_hamiltonian,eq:per_edge_color,eq:per_gate} with different, random
initial parameters multiple times and also once with initial parameters
corresponding to a discretised annealing schedule.

\subsection{Different Parametrisations}
\label{sec:different-parametrisations}

\begin{figure}[htb!]
  \centering
  \input{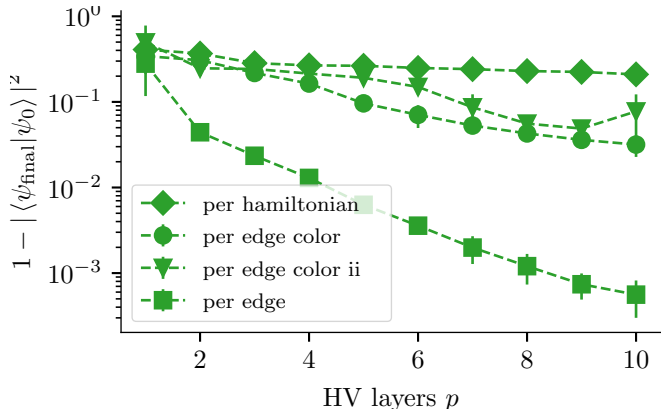}
  \caption{
    Infidelity with the ground state after parameter optimisation  as a
    function of ansatz layers $p$ for the four different  ansatz circuits on
    the $2 \times 8$ lattice.  Results are shown for 20 runs per data point
    with the initial parameters chosen uniformly random within
    $[0, \frac{1}{p}]$. The error bars reflect the standard deviation  between
    the 20 different runs.
  }%
  \label{fig:parametrisation_comparison}
\end{figure}

\Cref{fig:parametrisation_comparison} shows the difference between the three ansatz circuits
\cref{eq:per_hamiltonian,eq:per_edge_color,eq:per_gate} and a modified ``per
edge color ii'' ansatz circuit where the $e^{-i \theta_{i,0} H_0}$ was omitted
on the 2×8 lattice. The infidelity  with the true ground state
decays exponentially as a function of ansatz layers for the first 
three ansatz circuits. As expected, the more expressive ``per edge'' ansatz
represents the ground state better at lower depths than
the ``per edge color'' or ``per hamiltonian'' ansätze. The ``per edge color ii''
ansatz saves one layer of 2-qubit gates per ansatz layer, but at the cost 
of significantly worse fidelities, compared with the ``per edge color'' ansatz.
This remained true for other lattice sizes and hence we omit the results for this 
ansatz circuit from now on.

\begin{figure*}[htb!]
  \centering
  \input{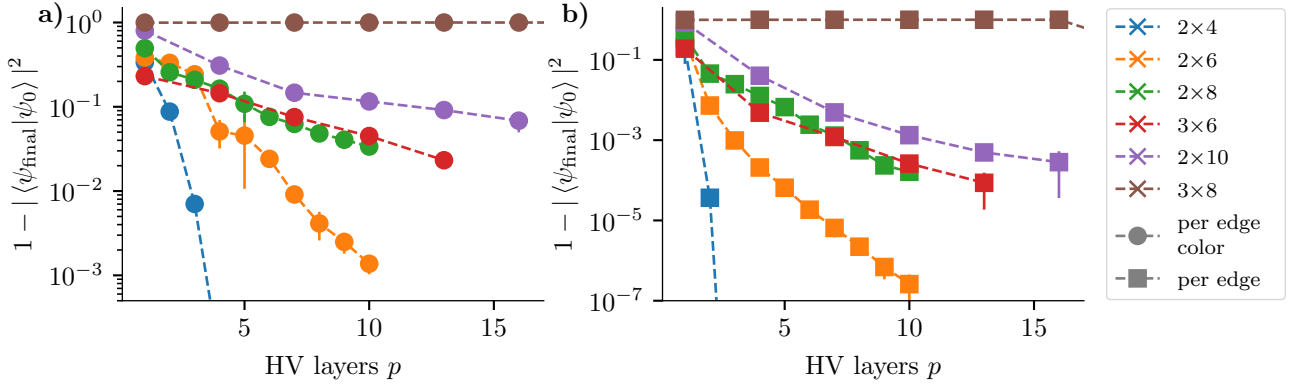}
  \caption{
    Infidelity with the ground state after parameter optimisation  as
    a function of ansatz layers $p$ for different lattice sizes. \textbf{a)} 
    the ``per edge color'' ansatz with one parameter per layer. \textbf{b)} the more 
    expressive ``per edge'' ansatz with one parameter for each gate.
    Results are shown for 10, 10, 4, 2, 2, 2 and 2 runs per data point for 
    the $2\times 4$, $2\times 6$, $2\times 8$, $3\times 6$, $2\times 10$ and
    $3 \times 8$ lattice, respectively. As for \cref{fig:parametrisation_comparison}
    the initial parameters were chosen uniformly random within $[0, \frac{1}{p}]$
    and the error bars reflect the standard deviation  between the different runs.
    The $2 \times 4$ data is not fully shown because $p=5$ was enough to produce
    the ground state with an infidelity of only $10^{-15}$, the machine precision.
  }%
  \label{fig:lattice_comparison}
\end{figure*}

\Cref{fig:lattice_comparison} compares the expressiveness of the ansatz
circuits between the different lattice sizes for the ``per edge color'' ansatz \textbf{a)}
and the ``per edge'' ansatz in \textbf{b)}. Generally, the ground state of
smaller lattices is better represented by shallow circuits than for larger
lattices. In fact, for the 3×8 lattice 16 ansatz layers were not enough to achieve
substantial overlap with the ground state even for the most expressive ansatz
circuits. This is due to the fact that for this particular lattice 
the gap between the ground state and the first excited state is very small 
compared to the energy. More details are found in \cref{sec:the_3x8_lattice}.
Except for the $3 \times 8$ lattice, the exponential decay of the infidelity as
a function of $p$ holds still true.

\subsection{Required circuit depths}
\label{sec:required-circuit-depths}

\begin{figure}[H]
  \centering
  \input{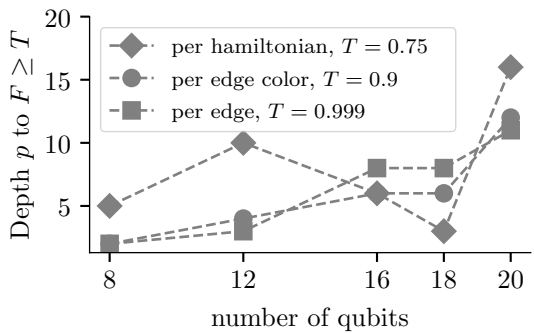}
  \caption{
    Number of ansatz layers $p$ needed to represent the ground state with 
    a threshold fidelity $T$ as a function of qubits for the three different 
    ansatz circuits and different thresholds.
    For the $3 \times 6$ and $2 \times 10$ lattices we ran the simulations only 
    for $p = 1 , 4, 7, \ldots $; The missing intermediate points are obtained via
    linear interpolation from the data shown in \cref{fig:lattice_comparison}.
    The $3\times 8$ data is not shown here because significant fidelity with 
    the ground state was only attainable with deep circuits and an informed 
    choice of initial parameters. For more details see \cref{sec:the_3x8_lattice}.
  }%
  \label{fig:depth_for_fidelity}
\end{figure}

In \cref{fig:lattice_comparison} we saw that larger lattices needed deeper
ansatz circuits to accurately represent the ground state. This is made more
quantitative in \cref{fig:depth_for_fidelity}, where we show the depth required
to reach a given fidelity as a function of qubits.
For the $2 \times x$ lattices this data is consistent with the needed depth 
scaling linearly with the lattice diameter. Because VQE for the $3 \times 8$
lattice failed to converge to the ground state we do not have enough data 
to extrapolate the scaling of the required depth for lattices with equal length 
in both dimensions. But if we assume that entanglement spreads linearly with 
the circuit depth then a scaling as low as $\sqrt{n_{qb}}$ seems 
possible. Still, given that we obtained high fidelity for 
only five different lattices, \cref{fig:depth_for_fidelity} should be
interpreted with caution.

\subsection{Barren plateaus?}
\label{sec:barren-plateaus}

\begin{figure}[H]
  \centering
  \input{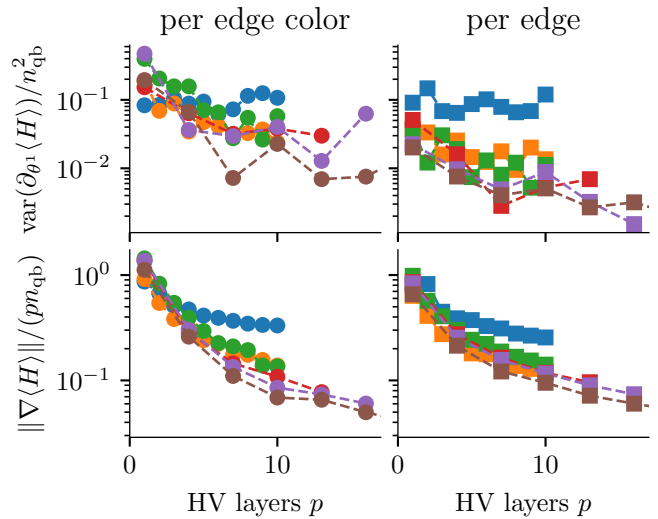}
  \caption{
    The variance of the first component of the gradient (upper row) and norm of
    the whole gradient (lower row) as a function of the number of ansatz layers
    $p$ for the ``per edge color'' (left column) and ``per edge'' (right column)
    parametrisation for different lattices. In all cases, we rescaled with the
    number of qubits $n_{\mathrm{qb}}$, because the number of terms in the
    Hamiltonian scales with the number of qubits. For the norm of the whole
    gradient we also rescaled with $1/p$, because the number of parameters
    scales with $p$.  The colour and marker shape coding is the same as
    elsewhere in this paper (e.g. \cref{fig:lattice_comparison}).
  }%
  \label{fig:ansatz_comparison_gradients}
\end{figure}

The results in \cite{Wiersema_2020} indicate that the HV ansatz exhibits only
mild barren plateaus in the case of the XXZ and transverse field Ising model
(TFIM) on 1D chains. To investigate whether the same is true for our ansatz
circuits for the KAFH, we evaluated the cost function gradients at the first
five points during each optimisation run, i.e. at essentially uniformly random
points within $[0, 1/p]^{n_{\mathrm{params}}}$ and long before the optimisation
had converged. The results for the ``per edge'' and the ``per edge color''
parametrisation are shown in \cref{fig:ansatz_comparison_gradients}.

We see in the upper row of \cref{fig:ansatz_comparison_gradients} that 
the variance of the first component of the gradient of the cost function does
decay as a function of circuit depth. From the figures it is not clear if 
it is indeed an exponential decay, but the decay rate appears to lie 
between the fast decay reported by McClean et al.~\cite{McClean_2018} for 
the random circuits and the much slower decay reported by Wiersema et al. for the
TFIM. To understand the scaling of the magnitude of the gradient as a 
function of the number of qubits, we would need to consider much deeper 
circuits (cf.\ fig.\ 4 of \cite{McClean_2018}) that were out of the scope of 
this work. It is also important to note that in the case of the ``per edge''
parametrisation the number of parameters scales as
$n_{\mathrm{params}} \sim p n_{\mathrm{qb}}$  compared with the other 
two parametrisations, where it scales as $n_{\mathrm{params}} \sim p$. This 
means that each individual entry of the gradient in the lower right pane
of \cref{fig:ansatz_comparison_gradients} is much smaller than in the 
lower left pane.

\subsection{Observables}
\label{sec:results-observables}

\begin{figure}[htb!]
  \centering
  \input{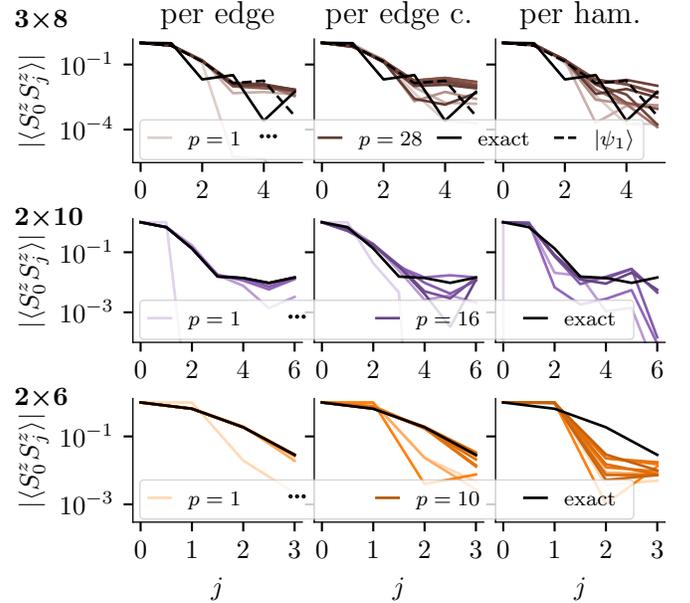}
  \caption{
    Spin-spin correlation along the straight lines marked with bigger nodes in
    the $2 \times 6$, $2 \times 10$ and $3 \times 8$ lattice in
    \cref{fig:graphs} for the state found by VQE with different ansatz depths
    (coloured) and for the exact ground state (black) as well as the first
    excited state (dashed), for the $3\times 8$ lattice.
  }%
  \label{fig:spin-spin-correlation}
\end{figure}

\begin{figure}[htb!]
  \centering
  \input{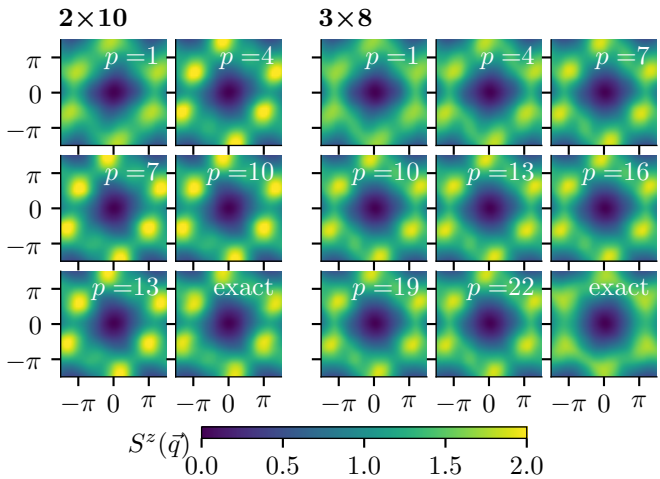}
  \caption{
    Comparison of the static structure factors
    $S^z(\vec{q}) = \frac{1}{N} \sum_{\vec{i},\vec{j}} e^{\imath \vec{q} \cdot (\vec{i} - \vec{j})} \braket{S^z_i S^z_j}$
    of the VQE states for different ansatz depths $p$ and the exact ground
    state for the $2\times 10$ and the $3 \times 8$ lattice with the ``per edge
    color'' parametrisation. $\vec{q}$ is measured in units of the inverse
    lattice spacing, hence we show a little more than the first Brillouin zone
    here to make the periodicity more more clear. As in
    \cref{fig:spin-spin-correlation} the $2 \times 10$ and the $3 \times 8$
    lattice showed the biggest difference between the VQE state and the exact
    ground state.
  }%
  \label{fig:static-structure-factor}
\end{figure}

The results in \cref{fig:lattice_comparison,fig:depth_for_fidelity,fig:large_lattices_fidelities} 
indicate that fairly deep circuits with many parameters are needed to obtain 
fidelities larger than 99.9\% for larger lattices. However, the fidelity is a 
\emph{global} figure of merit that gives error bounds on \emph{all} observables,
not only on \emph{local} ones. And ultimately, the goal of the experiments will
be to measure mostly local observables, like those mentioned on
\cref{sec:probing-observables}. 
To understand how well the states found by VQE reproduce local 
observables of the exact ground state, we compare the spin-spin correlation 
function and the static structure factor in the VQE states for different 
ansatz depths with the exact spin-spin correlation function and static 
structure factor of the exact ground state.
The results are shown in 
\cref{fig:spin-spin-correlation,fig:static-structure-factor}.

As expected, the more expressive ansätze or deeper circuits reproduce the exact
ground state spin-spin correlations better. Reassuringly, the spin-spin
correlations for $0 \leq j \leq 3$ of the VQE states for the $2 \times 6$ and
the $2 \times 10$ lattice match the exact spin-spin correlations equally well,
even though the global fidelity of the $2 \times 6$ VQE states with the exact
ground state is much better than that of the $2 \times 10$ lattice (see
\cref{fig:lattice_comparison}). We found the same to be true for the $2 \times
8$ and the $3 \times 6$ lattice. This implies that even though it is hard to
reach high fidelity with the exact ground state for large lattices the VQE
states may still represent local observables well. For the $3 \times 8$ lattice,
on the other hand, there is a marked discrepancy between the spin-spin correlations
of the VQE states and of the exact ground state, presumably because in this 
case the VQE state has a large fidelity with the first excited state and 
not with the ground state (cf.\ \cref{fig:3x8_fidelities}).

The VQE states reproduce the static structure factors of the ground state
closely, as can be seen from \cref{fig:static-structure-factor}. On larger
lattices, we expect the peaks of the static structure factor to not lie on the
corners of the first Brillouin zone, like they do here, but within the first
Brillouin zone, if the periodicity of the ground state is larger than one unit
cell, as it is for the ground state conjectured in~\cite{Nikolic_2003}.  It is
important to note, that the locations of the peaks
in~\cref{fig:static-structure-factor} are simply at the locations of the
Fourier transformed lattice. Instead, the features that indicate closeness
between the VQE state and the exact ground state are the shapes of the peaks
and the secondary peaks between the main peaks.

\section{Discussion and Outlook}%
\label{sec:discussion_and_outlook}

In this work we investigated the performance of the VQE with the HV ansatz on
the KAFH for different lattice sizes numerically. We also considered whether
the fidelity of the states produced by VQE with the exact ground state is a too
restrictive figure of merit by comparing different local, physically
interesting observables in the VQE states and the exact ground state.
Furthermore, we calculated the cost function gradients at random points in the
parameter space to study how severe the barren plateau problem is with our
ansatz.

In accordance with the results in \cite{Cade_2019,Biamonte_2020}, we find again 
that the relative energy error and infidelity with the ground state decay 
exponentially as a function of ansatz depth. We also find, again in accordance 
with above results, that the number of layers needed to get to a fixed fidelity
grows with the system size (see \cref{fig:depth_for_fidelity}) and 
that in the presence of small energy gaps VQE may fail to find the ground state
altogether (see \cref{fig:3x8_fidelities}).
Furthermore, the results in \cref{sec:results-observables} show that even at
lower fidelities the states found with VQE represent the values of local
observables well. If this remains true for larger lattices and other local
observables, this hints that that requiring to find the ground state with a
given fidelity is too strict a requirement, if one is mainly interested in the
values of local observables.

Even though the barren plateaus of our ansatz are not as pronounced as for the 
hardware efficient ansatz \cite{McClean_2018}, the gradients still decay as a 
function of circuit depth (and presumably also as a function of system size).
This is particularly true for the more expressive ansatz circuits with 
more free parameters. On real hardware, if the cost function is estimated from 
noisy samples, this opens up the question of the best classical optimisation
algorithm to use within VQE.  To the best of our knowledge, there exist no
classical optimisation algorithms that are known to perform well for noisy,
derivative-free problems with hundreds of parameters.

With the depth of 228 two-qubit gates reported in \cite{Arute_2020} as a
circuit depth budget and $\sim 50$ qubits as a system size that is not feasible
to simulate using exact, classical
methods~\cite{Sindzingre_2009,Laeuchli_2019} useful quantum advantage
seems in reach with VQE for the KAFH. It should be noted, nevertheless, that 
approximate methods like DMRG~\cite{Jiang_2008,Yan_2011,Schollwoeck_2012}
have been used to investigate the ground state properties of the KAFH 
for systems with hundreds of sites, while other tensor network based
approaches claim to handle thousands~\cite{Mei_2017} of sites or even 
give infinite size results~\cite{Liao_2017}. However, most tensor network based 
approaches are biased towards low-entanglement solutions and they all have,
by design, limited expressibility. Therefore, we hope that VQE can serve as 
an intermediate tool between exact diagonalisation, which is limited to small
system sizes, but gives exact results for these and can express arbitrary
states, and tensor network methods, which work for much larger systems, but 
have limited expressibility and known biases towards certain solutions.

To get close to the ground state with relatively shallow circuits, one will 
almost certainly need parametrisations with one parameter per gate which 
yields hundreds of parameters to be optimised. Previous VQE experiments have 
not considered cost functions with that many free parameters and hence the best 
classical optimisation algorithm for such problems is yet to be found.

Computing properties of the ground state accurately will require the use of
error-mitigation techniques or small-scale quantum error-correction. As
discussed above, the symmetries of the KAFH allow a simple notion of error
detection, by checking the total spin in each direction. Many other techniques
targeted at NISQ-era quantum computers are now known~\cite{endo21}.

On the $3 \times 8$ lattice we found that VQE consistently finds instead of the
ground state the first excited state, which happens to be closer to the dimer
covering state that we chose as an initial state. This hints that choosing an
initial state that is already close to the ground state (if such a state is
known and easily preparable) can help the performance of VQE. Kattemölle and
van Wezel~\cite{Kattemoelle_2021_personal} observed similar behaviour for a
periodic lattice with 6 spins.  However, minimizing the infidelity instead of
energy the recovered the exponential decay of the infidelity with $p$ observed
on all other lattices. Although measuring the fidelity with the true ground
state is not possible in the real experiment, this still shows that the
circuits are expressive enough to represent the true ground state with high
fidelity, although it may be hard to find if the gap to the first excited state
is only small.

Independently, Kattemölle and van Wezel~\cite{Kattemoelle_2021} also studied VQE
with the HV ansatz on the KAFH, albeit with a different ansatz circuit and
initial state.  Unlike ours, their ansatz circuit does not break the
translational symmetry of a infinite lattice. Still, their initial state
necessarily does, because there exists no dimer covering of the kagome lattice
that does not break the translational symmetry.  They report slightly better
scaling of the infidelity as a function of $p$ for a 20 site lattice.  This
hints at the possibility that the performance of VQE with the HV ansatz also
depends on the exact ansatz circuits and one might want to experiment with
different mappings of the kagome lattice onto the square lattice of the
hardware.

\begin{acknowledgments}
  The authors would like to thank Filippo Gambetta, Raul Santos and other
  members of the Phasecraft team for helpful discussions and feedback on the
  early drafts of this paper.
  This project has received funding from the European Research Council (ERC)
  under the European Union's Horizon 2020 research and innovation programme
  (grant agreement No.\ 817581) and from the EPSRC grant EP/S516090/1. Google
  Cloud credits were provided by Google via the EPSRC Prosperity Partnership in
  Quantum Software for Modeling and Simulation (EP/S005021/1).  All data is
  available at the University of Bristol data repository
\cite{Bosse2021_data}.
\end{acknowledgments}

\bibliography{paper}

\appendix

\section{Additional results on large lattices}
\label{sec:appendix}

\subsection{The \texorpdfstring{$3 \times 8$}{3x8} lattice}
\label{sec:the_3x8_lattice}

\begin{figure}[ht]
  \centering
  \input{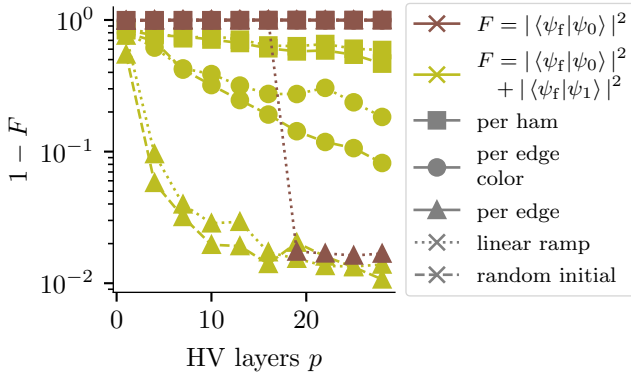}
  \caption{
    Infidelity with the ground state and the ground state + first excited state 
    as a function of ansatz layers $p$ for the three different ansatz circuits 
    on the $3\times 8$ lattice. The ``random'' initial parameters are, again, 
    chosen uniformly random within $[0, \frac{1}{p}]$ while the ``linear ramp''
    initial parameters are chosen according to a discretised, trotterised 
    annealing schedule.
  }%
  \label{fig:3x8_fidelities}
\end{figure}

\begin{figure}[ht]
  \centering
  \input{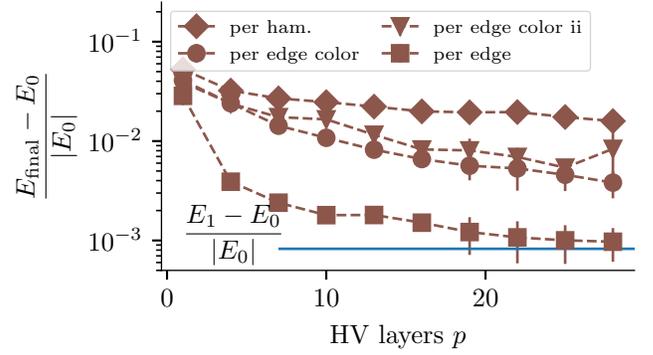}
  \caption{Scaling of the relative energy error as a function of $p$ for the
    $3 \times 8$ lattice with different ansatz circuits. Results are shown for 3
    runs per data point and with the initial parameters chosen uniformly random
    within $[0, \frac{1}{p}]$.  The error bars reflect the standard deviation
  between the different runs.}%
  \label{fig:3x8_energies}
\end{figure}

VQE was only able to find the ground state of the $3 \times 8$ lattice with the
most expressive ``per edge'' ansatz with $p \geq 19$ ansatz layers and using
initial parameters corresponding to a trotterised annealing schedule. However,
changing the $y$-axis in \cref{fig:lattice_comparison} from infidelity to
relative energy error $\frac{E_{\mathrm{final}} - E_0}{|E_0|}$ we observed the
same exponential decay for the $3 \times 8$ lattice as for the other lattices,
until the energy of the first excited state $E_1$ is reached, see
\cref{fig:3x8_energies}. The reason is the very small relative energy gap
$\frac{E_1 - E_0}{|E_0|} \approx 0.0008$ between the ground state and the first
excited state. And in fact, when considering not only the projection onto the
ground state, but also onto the first excited state we get much better
fidelities, as is shown in \cref{fig:3x8_fidelities}. Moreover, we found that
the first excited state is close to a dimer covering that shares many dimers
with the dimer covering that we use as an initial state.

\subsection{Other large lattices}
\label{sec:other_larger_lattices}

\begin{figure}[bh]
  \centering
  \input{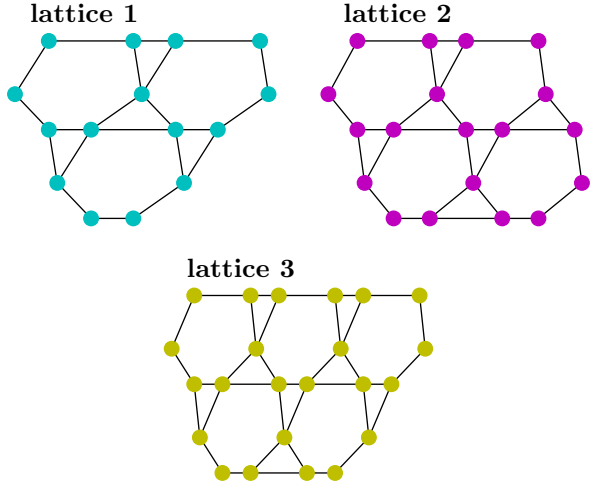}
  \caption{
    Three lattices with 1, 2, and 3 completely enclosed triangles on 
    15, 19 and 23 sites, respectively.
  }%
  \label{fig:large_graphs}
\end{figure}

\begin{figure}[htb]
  \centering
  \input{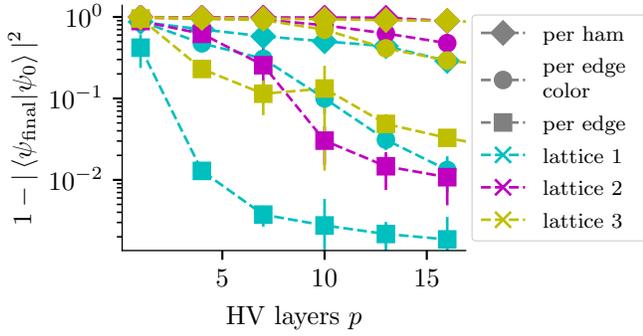}
  \caption{
    Infidelity with the ground state as a function of ansatz layers $p$ for the
    three different ansatz circuits on the lattices shown in
    \cref{fig:large_graphs}. As before, the initial parameters are chosen
    uniformly random within $[0, \frac{1}{p}]$.
  }%
  \label{fig:large_lattices_fidelities}
\end{figure}

Outside of the six lattices shown in \cref{fig:kagome_lattice_square_list} that
fit into a rectangular part of a square lattice, we also considered three other
large lattices shown in \cref{fig:large_graphs}. Since they all have an odd
number of sites $n_{\mathrm{qb}}$, we chose an initial state in the $S^z =
-\frac{1}{2}$ sector and patched $H_0$ such that it acts on
$\frac{n_{\mathrm{qb}} - 1}{2}$ edges and hence has its ground states in the
$S^z = \pm \frac{1}{2}$ sectors.  We ran the same simulations as for
\cref{fig:lattice_comparison} for these three graphs as well and show the
results in \cref{fig:large_lattices_fidelities}.

For lattice 2 and lattice 3 we see the same exponential decay in infidelity 
as before for all three parametrisations. But for lattice 1 and the 
``per edge'' parametrisation the infidelity decays much slower for 
$p \geq 7$, similar to what we see in \cref{fig:3x8_fidelities} for 
$p \geq 10$.

\section{An edge 4-colouring of the kagome lattice}
\label{sec:a-4-colouring-of-the-kagome-lattice}

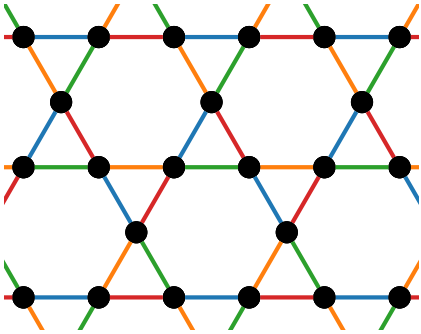
\begin{figure}[h]
  \centering
\newcommand{\kagomeunitcell}[5]{
  \begin{scope}[shift={#1},yscale=#2]
    \node [draw, shape=circle, fill=black, inner sep=0pt, minimum size=8pt] (0) at (0,0){};
    \node [draw, shape=circle, fill=black, inner sep=0pt, minimum size=8pt] (1) at (1,0){};
    \node [draw, shape=circle, fill=black, inner sep=0pt, minimum size=8pt] (2) at (1.5,{.5*sqrt(3)}){};
    \node [draw, shape=circle, fill=black, inner sep=0pt, minimum size=8pt] (3) at (2,0){};
    \draw[color=#3, ultra thick] (0) -- (1);
    \draw[color=#4, ultra thick] (1) -- (2);
    \draw[color=#5, ultra thick] (2) -- (3);
  \end{scope}
}

\begin{tikzpicture}
  \clip (1.75,{-2.25 * sqrt(3)}) rectangle (7.25,{0.25*sqrt(3)});

  \foreach \x in {0, 2, 4, 6}{
    \kagomeunitcell{(\x,0)}{1}{c0}{c1}{c2}
  }
  \foreach \x in {1, 3, 5, 7}{
    \kagomeunitcell{(\x,0)}{-1}{c3}{c1}{c2}
  }
  \foreach \x in {1, 3, 5, 7}{
    \kagomeunitcell{(\x,{-sqrt(3)})}{1}{c1}{c0}{c3}
  }
  \foreach \x in {0, 2, 4, 6}{
    \kagomeunitcell{(\x,{-sqrt(3)})}{-1}{c2}{c0}{c3}
  }
  \foreach \x in {0, 2, 4, 6}{
    \kagomeunitcell{(\x,{-2*sqrt(3)})}{1}{c0}{c1}{c2}
  }
  \foreach \x in {1, 3, 5, 7}{
    \kagomeunitcell{(\x,{-2*sqrt(3)})}{-1}{c3}{c1}{c2}
  }

\end{tikzpicture}
  \caption{An edge 4-colouring of the kagome lattice}
  \label{fig:kagome_lattice_coloring}
\end{figure}

\end{document}